\documentclass{elsart}

\usepackage{natbib}

 \usepackage{graphicx}

\usepackage{amssymb}

\begin{document}

\begin{frontmatter}



\title{VHE Gamma-ray supernova remnants}


\author{S.~Funk\thanksref{sf}}
\ead{Stefan.Funk@slac.stanford.edu}
\address{Kavli Institute for Particle Astrophysics and Cosmology,
  SLAC, 2575 Sand Hill Road, PO Box 0029, Stanford, CA-94025, USA}
\thanks[sf]{previously at: Max-Planck-Institut f\"ur Kernphysik,
  P.O. Box 103980, D 69029, Heidelberg, Germany}

\begin{abstract}
Increasing observational evidence gathered especially in X-rays and
$\gamma$-rays during the course of the last few years support the
notion that Supernova remnants (SNRs) are Galactic particle
accelerators up to energies close to the ``knee'' in the energy
spectrum of Cosmic rays. This review summarises the current status of
$\gamma$-ray observations of SNRs. Shell-type as well as plerionic
type SNRs are addressed and prospect for observations of these two
source classes with the upcoming GLAST satellite in the energy regime
above 100~MeV are given.
\end{abstract}

\begin{keyword}
\sep Supernova remnant
\sep Pulsar Wind Nebula
\sep Gamma-ray astronomy
\PACS 98.38.Mz	\sep Supernova remnants
\PACS 98.70.Rz	\sep $\gamma$-ray sources
\PACS 98.70.Sa	\sep Cosmic rays (origin, acceleration, and interactions)

\end{keyword}

\end{frontmatter}

\section{Introduction}
\label{sec::Intro}
Supernova remnants (i.e. remnants of Supernova explosions) are
commonly considered to be Cosmic particle accelerators. In this review
paper I will summarise experimental evidence, gathered through
$\gamma$-ray observations mainly in the VHE regime supporting this
notion (please note, that in the following $\gamma$-ray will be used
to stand for VHE $\gamma$-rays). Typically, SNRs were detected through
radio observations~\citep{Green}. Recent advances in the understanding
of these objects has been made through X-ray observations with
instruments such as ASCA, BeppoSax, XMM-Newton and Chandra and through
$\gamma$-ray observations with instruments such as HEGRA (High Energy
Gamma Ray Astronomy), and H.E.S.S.\ (High energy stereoscopic
system). Based on morphological properties, SNRs can be classified
into three broad categories: {\emph{shell-like}}, {\emph{plerionic}}
(also called Pulsar Wind Nebulae or Crab-like) connected to a Pulsar
and {\emph{composite}} (in which both, a shell and a Plerion are
present), the later often showing markedly different radio and X-ray
morphologies. For an excellent detailed review on plerionic SNRs, see
e.g.~\citet{GaenslerReview}.

SNRs are thought to be responsible for the acceleration of Cosmic rays
up to energies around the ``knee'' ($\sim 10^{15}$~eV) at which the
spectrum of Cosmic rays significantly hardens from $\sim 2.7$ to $\sim
3.2$. This statement is backed by experimental facts, as well as by
theoretical considerations. {\emph{Experimental evidence}} is lent
mainly by ({\bf 1}) X-ray observations of young shell-type SNRs such
as SN\,1006~\citep{SN1006ASCA}, and Cas~A~\citep{BeppoSaxCasA}, in
which sites dominated by hard non-thermal {\emph{X-ray synchrotron}}
emission were found, indicating an electron population extending up to
$\sim100$~TeV, far beyond thermal energies. ({\bf 2}) {\emph{Very high
energy (VHE) $\gamma$-ray}} observations ($> 100$~GeV) revealed sites
of non-thermal particle populations~\citep{HESSRXJ1713,
HESSRXJ1713_II, HESSRXJ1713_III}. Through theoretical considerations
is has been known for a long time that Supernova explosions release
just about the right amount of energy into their surrounding to
account for the energy budget of the Cosmic rays (assuming that they
convert $\sim 10\%$ of their energy into kinetic energy of the Cosmic
rays)~\citep{Ginzburg}. Furthermore well-established theoretical
models exists explaining how particles can be accelerated in Supernova
shock waves to energies approaching the {\emph{knee}}. In shell-type
SNRs particles are accelerated in the expanding shock waves through
diffusive shock (also called first order Fermi)
acceleration~\citep{Bell, BlandfordOstriker, Drury, Blandford, Jones,
MalkovDrury}. In plerionic SNRs particles are accelerated to
non-thermal energies in the termination shock between the relativistic
outflow of electrons from the pulsar surface and the outer
nebula. Predictions on the $\gamma$-ray visibility of SNRs (later
confirmed by H.E.S.S.\ $\gamma$-ray observations although there is
still an experimental ambiguity in the underlying particle population
resonsible for the $\gamma$-ray emission) were given
by~\citet{DrurySNRs}.

Since charged particles below the knee at $10^{15}$~eV are deflected
in ubiquitous magnetic fields on their way from the origin to us, we
have to turn to neutral messengers to reveal the acceleration sites
(the gyroradius of 1~TeV cosmic rays in a magnetic field of
$\mu$G-scale is of the order of 0.1~pc, much smaller than the
thickness of the Galaxy of 200--300~pc). Since neutrino detectors have
not yet proved to be sensitive enough to detect neutrinos from
astrophysical sources (apart from the Sun and the direct Supernova
explosion SN~1987A), observations in the radio, X-rays and
$\gamma$-ray wavebands are so far our best access to non-thermal
acceleration processes in SNRs. $\gamma$-rays (and neutrinos) are
produced in hadronic interactions with subsequent pionic decay and can
reveal the acceleration sites since they travel un-deflected from
their origin. However, $\gamma$-rays not only reveal the sites of
hadronic acceleration; they also act as a tracer for energetic
electrons that produce $\gamma$-rays via IC scattering off background
photon fields (such as star-light or the Cosmic microwave background
(CMBR)). An ambiguity or duality therefore exists in the responsible
radiating particle population in most cases when detecting
$\gamma$-rays from astrophysical objects.

In spite of this ambiguity the detection of $\gamma$-rays above $\sim$
1~GeV from SNRs gives us direct access to particle acceleration
processes and the advantage of $\gamma$-rays in comparison to other
wavebands is that these are not affected by dust obscuration, which is
particularly important for the population of SNRs located within the
Galactic plane. A large volumne of the Galaxy can thus be probed for
$\gamma$-ray emission from SNRs by observations through the Galactic
disk. If SNRs are indeed sites of particle acceleration, $\gamma$-ray
emission is expected and one of the puzzling aspects of previous
$\gamma$-ray observations of SNRs was the rather low level of emission
compared to model predictions~\citep{Buckley, HillasReview}.

The history of soft $\gamma$-ray (or hard X-ray) detection of SNRs
started with the detection of the Crab Nebula in 1964 with a
scintillation counter detector flown on a balloon launched from
Palestine, Texas~\citep{XraysCrab1964}. Today hard X-rays up to
100~keV have been detected from various SNRs both young shell-types
such as Cas~A~\citep{BeppoSaxCasA, BeppoSaxCasAII} and
SN~1006~\citep{RXTESN1006, BeppoSaxSN1006} and plerionic-types such as
the Vela-X PWN~\citep{BeppoSaxVelaPWN} and
MSH~15-5{\emph2}~\citep{ASCAMSH, RXTEMSH, BeppoSaxMSH1552,
IntegralMSH}. Thin X-ray filaments in young shell-type SNRs detected
with high-angular resolution instruments such as XMM-Newton and
Chandra point to regions with high-magnetic fields (up to 0.5 mG) in
which electrons rapidly lose energy through synchrotron
emission~\citep{ASCARXJ1713, ASCAVelaJr, ChandraSN1006, XMM1713}. In
higher energies $\gamma$-rays COMPTEL detected the radioactive
$^{44}$Ti-line at 1.157~MeV from the two shell-type SNRs
Cas~A~\citep{ComptelCasA} and RX\,J0852.0--4622 (Vela
Junior)~\citep{ComptelVelaJr}. However, it should be noted, that
higher sensitivity INTEGRAL observations provided a confirmation for
this detection for Cas~A, but could not detect this $^{44}$Ti-line in
RX\,J0852.0--4622. Therefore, these claims are still somewhat
controversial.

EGRET at energies above 100~MeV did not detect prominent young
shell-type SNRs such as Tycho, Kepler, Cas~A or SN~1006, noticed
however several intriguing spatial coincidences of unidentified
sources in the Galactic plane with individual prominent radio SNRs,
such as W\,28, and $\gamma$-Cygni~\citep{Dermer, Esposito,
Romero}. The combination of source confusion especially in the
Galactic plane, caused by the rather poor angular resolution of the
EGRET instrument and the ambiguity in existing counterparts prevented
individual identifications. However, a statistical assessment shows a
4-5$\sigma$ effect when trying to correlate the population of EGRET
unidentified sources with the population of radio
SNRs~\citep{Dermer}. Also plerionic SNRs have not been unambiguously
identified with EGRET sources, although again intriguing associations
of EGRET unidentified sources with prominent plerions such as
PSR\,B1706--44, and the Kookaburra complex
exist~\citep{EGRETPWN}. From a population point-of-view PWN are one of
the best candidates to account for low-latitude slowly varying
unidentified EGRET sources as proposed by
\citet{RobertsEGRET}. Lately, using VHE $\gamma$-ray source positions
in the Kookaburra region~\citep{HESSKooka} the re-analysis of EGRET
data provided strong evidence of correlation of the PWN detected in
this region with the confused unidentified EGRET source
3EG\,J1420--6038~\citep{ReimerKookaburra}. This new approach might
prove a useful template for connection future GLAST and VHE
$\gamma$-ray detections. All these possible associations of source
classes with unidentified EGRET sources will hopefully be tested
following the launch of the upcoming GLAST satellite in late 2007.

The history of VHE $\gamma$-ray (E$> 100~$GeV) detections of SNRs
started again with the detection of the Crab Nebula, the first object
to be reported in this waveband by the Whipple
collaboration~\citep{WhippleCrab}. Various claims of detections of
shell-type SNRs have been made before the advent of the H.E.S.S.\
telescope system. Cas~A was detected by HEGRA in a very deep ($\sim
200$ hours) exposure~\citep{HegraCasA}. Detections of
SN\,1006~\citep{Cangaroo_SN1006} and
RX\,J1713.7--3946~\citep{Cangaroo_RXJ1713} have been reported by the
CANGAROO collaboration. With the advent of the H.E.S.S.\ telescope
system, for the first time a number of Galactic SNRs, both shell-type
and plerionic in nature could be established. In the following I will
describe these populations of cosmic accelerators along with prospects
for Supernova remnant observations with the upcoming GLAST
satellite. The outline of this paper is as follows:
Section~\ref{sec::HESS_shell} provides a description of advances made
through the VHE $\gamma$-ray detections of shell-type Supernova
remnants, Section~\ref{sec::HESS_pwn} provides the corresponding
description for Pulsar Wind Nebulae. Section~\ref{sec::HESS_new}
describes Supernova remnants found in $\gamma$-rays and later
identified as Supernova remnants, while Section~\ref{sec::glast}
summarises prospects for Supernova remnant observations with the
upcoming GLAST satellite.

\section{Gamma-ray observations of shell-type SNRs}
\label{sec::HESS_shell}
By means of data taken with the H.E.S.S.\ telescope system during the
first few years of operation for the first time in VHE $\gamma$-ray
astronomy resolved images of shell-type SNRs above 100~GeV could be
taken. In particular the SNRs RX\,J1713.7--3946 and RX\,J0852.0--4622,
with diameters of $\sim 1^{\circ}$ and $\sim 2^{\circ}$ respectively
could be resolved with unprecedented detail in this energy band. On
the other hand, SN\,1006, one of the SNRs most expected to emit
$\gamma$-rays in the energy band (due to the strong non-thermal X-ray
emission from the rims) was not detected in deep H.E.S.S.\
observations~\citep{HESSSN1006}. The upper limit derived by this
observations turned out to be an order of magnitude below the
previously reported CANGAROO detection. A reanalysis of the CANGAROO
data along with newer data from the CANGAROO-III detector is
consistent with the H.E.S.S.\ upper
limits~\citep{Cangaroo_SN1006_II}. Therefore in the following the
H.E.S.S.\ upper limits will be used in the discussion of $\gamma$-ray
emission from SN\,1006. Since several papers on both
RX\,J1713.7--3946~\citep{HESSRXJ1713, HESSRXJ1713_II, HESSRXJ1713_III}
and RX\,J0852.0--4622~\citep{HESSRXJ0852, HESSRXJ0852_II} have been
published by the H.E.S.S.\ collaboration, the main focus of this
review will lie on similarities and differences between the two
objects with the addition of comparisons to SN\,1006 as the most
prominent non-detected SNR where appropriate.

\begin{figure*}
  \centering
  \includegraphics[width=\textwidth]{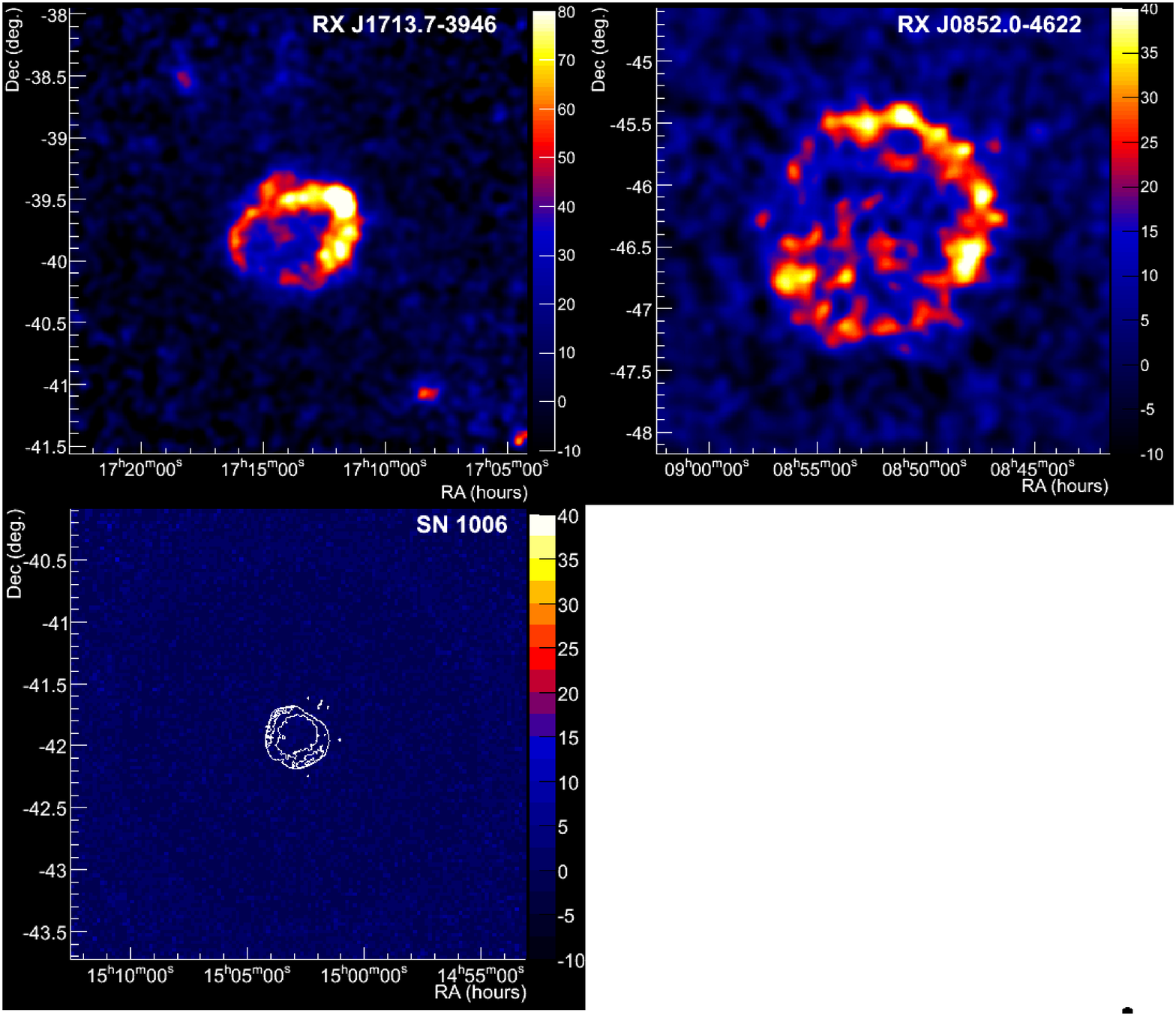}
  \caption{Acceptance-corrected smoothed excess maps of the
    $3.5^{\circ}\ \times 3.5^{\circ}$ fov surrounding the two
    prominent H.E.S.S.\ Supernova remnants RX\,J1713.7--3946 (2004 and
    2005 data)~\citep{HESSRXJ1713_III} and RX\,J0852.0--4622 (2005
    dataset)~\citep{MarianneBarcelona, HESSRXJ0852_II} and non-detected
    SN\,1006 (2004 dataset with VLA radio contours in
    white)~\citep{HESSSN1006}. The sky-regions shown are of similar
    size, indicating the large extent (2$^{\circ}$ diameter) of
    RX\,J0852.0--4622.  }
  \label{fig::shell_morphology}
\end{figure*}

Figure~\ref{fig::shell_morphology} shows $\gamma$-ray excess maps for
RX\,J1713.7--3946, RX\,J0852.0--4622, and SN\,1006. Both $\gamma$-ray
emitting objects show a shell-like structure with a surprising
resemblance of their respective X-ray morphology (the correlation
coefficients between $\gamma$-ray and X-ray counts are $\sim
60\%-80\%$). For both objects the X-ray emission is completely
dominated by non-thermal X-ray emission without traces of line
emission, exhibiting small filamentary structures that are interpreted
as zones where the magnetic field is high ($\sim 50\mu$G) such that
electrons rapidly lose energy through synchrotron emission in these
areas~\citep{XMM1713, Chandra1713, AschenbachVelaJr, XMMVelaJr}. Both
objects appear rather faint in radio with typical fluxes below or in
the several tenth of Jansky-regime for the whole shell, certainly
lower than what would be expected from equipartition
arguments~\citep{Lazendic}. The distance to both objects is somewhat
uncertain, for RX\,J1713.7--3946 it seems that a distance of $\sim
1$kpc is preferred from the column density inferred from X-ray
data. This distance would make RX\,J1713.7--3946 most likely the
remnant of the historical Supernova event of AD393. For
RX\,J0852.0--4622 distance estimates range from as close as the Vela
pulsar ($\sim 250$pc) to as far as the Vela Molecular Ridge ($\sim
1$kpc). The age ranges from $\sim 500$ years in the close case to
$\sim 5000$ years in the far case. Morphologically their $\gamma$-ray
emission, in particular the width of the shells is rather
different. The apparent width of the shell for RX\,J1713.7--3946
comprises 45\% of the radius of the SNR, while the for
RX\,J0852.0--4622 it approximates to 20\% of the radius. There is no
apparent correlation between the dense molecular material surrounding
RX\,J1713.7--3946 as measured by the NANTEN telescope and the VHE
$\gamma$-ray emission as measured by H.E.S.S., but in fact, assuming a
typical energy of $1\times 10^{50}$ ergs in accelerated protons, the
density needed to explain the $\gamma$-ray flux through hadronic
interactions is only 1~cm$^{-3}$.

SN\,1006 is somewhat distinct in its multi-frequency picture in that
its surface brightness is higher in radio ($\sim 100$
Jy)~\citep{RadioSN1006, RadioSN1006_II} showing a pronounced
shell-like structure. The X-ray emission, especially in the shell is
dominated by non-thermal emission up to $\sim 10$keV. SN\,1006 was not
detected in a deep (1000~ksec) INTEGRAL exposure above
20~keV~\citep{IntegralSN1006} and was also not detected in sensitive
H.E.S.S.\ observations~\citep{HESSSN1006}. The density surrounding the
source was estimated from X-ray as well as optical observations and
values as low as n$= 0.05$ cm$^{-3}$ have been invoked to explain the
apparent absence of $\gamma$-ray emission in a hadronic scenario. From
the H.E.S.S.\ non-detection assuming a leptonic $\gamma$-ray emission
scenario on the CMBR a lower limit on the post-shock magnetic field of
$B > 25\mu$G can be derived~\citep{HESSSN1006}. Higher values of the
magnetic field in excess of 40$\mu$G have been derived from X-ray
observation and application of diffusive shock acceleration scenarios,
so the lower limit on the magnetic field is not in contradiction to
these values.

\begin{figure*}
  \centering
  \includegraphics[width=0.9\textwidth]{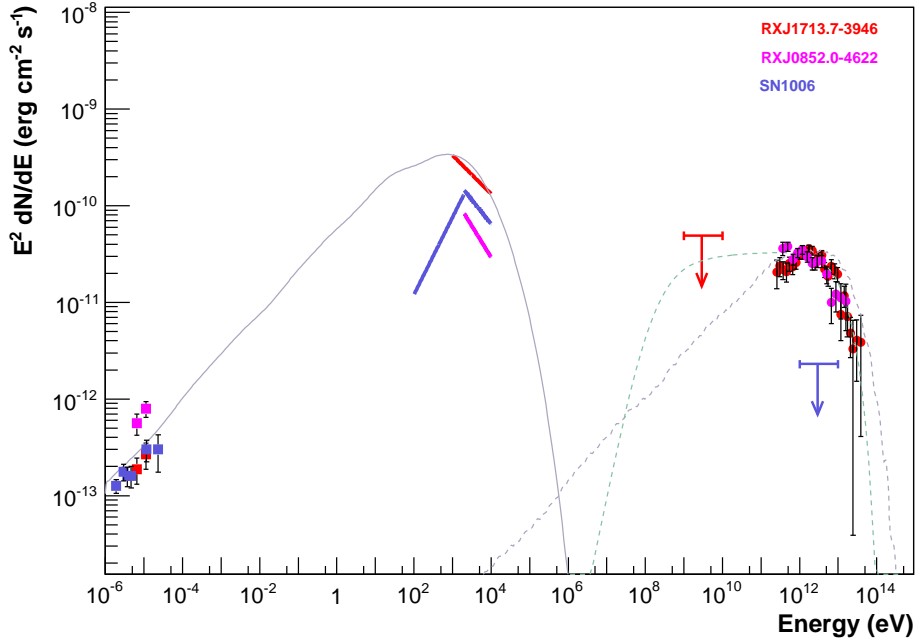}
  \caption{Spectral energy distribution for the Supernova remnants
  RX\,J1713.7--3946 (light red), RX\,J0852.0--4622 (pink), SN\,1006
  (blue). Also shown is a hadronic model (dotted green) and a
  time-dependent leptonic one-zone model (dashed grey: IC emission,
  solid grey: synchrotron emission). The parameters for this model
  are: B-Field: 9$\mu$G, age: 1.7 kYears, Electron photon index: 2.1,
  Electron Cutoff: 80 TeV.}
  \label{fig::shell_spectra}
\end{figure*}

Comparing the energy spectra of the two $\gamma$-ray detected SNRs,
strong similarities can be made out. The spectral energy distribution
(SED) for the three shell-type SNRs discussed here is shown in
Figure~\ref{fig::shell_spectra} along with model spectra, showing
typical leptonic and hadronic $\gamma$-ray emission models. As can be
seen from this plot RX\,J1713.7--3946 and RX\,J0852.0--4622 show a
remarkably similar $\gamma$-ray energy spectrum with a rather flat
$E^{-2}$-type distribution at lower energies with a deviation from
this power-law at higher energies. The flat spectrum at lower energies
has advocated claims that the $\gamma$-rays might be generated by
pionic decays rather than Inverse Compton
scattering~\citep{HESSRXJ1713_II, HESSRXJ1713_III}. However,
\citet{PorterMoskalenko} claim that the data can be well fitted in
terms of a leptonic model when applying an unbroken electron spectrum
along with the Galactic radiation fields.  Therefore, at the moment,
no strong conclusions can be drawn from the spectral shape on the
particle population responsible for the $\gamma$-ray emission. The
upcoming GLAST satellite, measuring in the energy range between 30~MeV
and 300~GeV might be able to distinguish between hadronic and leptonic
$\gamma$-ray production mechanisms. Also interesting to note is that
the H.E.S.S.\ $\gamma$-ray upper limit for SN\,1006 is more than an
order of magnitude below these $\gamma$-ray detections and therefore
starts to be rather constraining for the values of the magnetic field
(in a leptonic scenario) or the ambient matter density (in a hadronic
scenario).

These first unambiguous detections of individual shell-type SNRs
allowed for important advances in the understanding $\gamma$-ray
emission from these objects. However, the open question remains what
differentiates non-detected SNRs such as SN\,1006 from prominent
$\gamma$-ray emitters such as RX\,J1713.7--3946.

\section{Gamma-ray observations of Pulsar Wind Nebulae}
\label{sec::HESS_pwn}
Pulsar wind nebulae (PWN) or Plerions are objects powered by a
relativistic particle outflow (electrons and positrons) from a central
source -- a pulsar. This pulsar is a rapidly rotating neutron star
generated in the Supernova event. The wind of relativistic particles
flows freely out until the outflow pressure is balanced by that of the
surrounding medium. At that point a standing termination shock is
formed at which particles are accelerated~\citep{KennelCoroniti,
AhaAtoKif97}. The existence of electrons accelerated to energies
$>100$~TeV in such PWN has been established by X-ray observations of
synchrotron emission, e.g. in the Crab nebula~\citep{ChandraCrab}. VHE
$\gamma$-rays are generated in PWN from the high-energy electrons by
non-thermal bremsstrahlung or inverse Compton (IC) scattering on
photon target fields, such as the cosmic microwave background (CMBR)
or star-light.  

\begin{figure*}
  \centering
  \includegraphics[width=\textwidth]{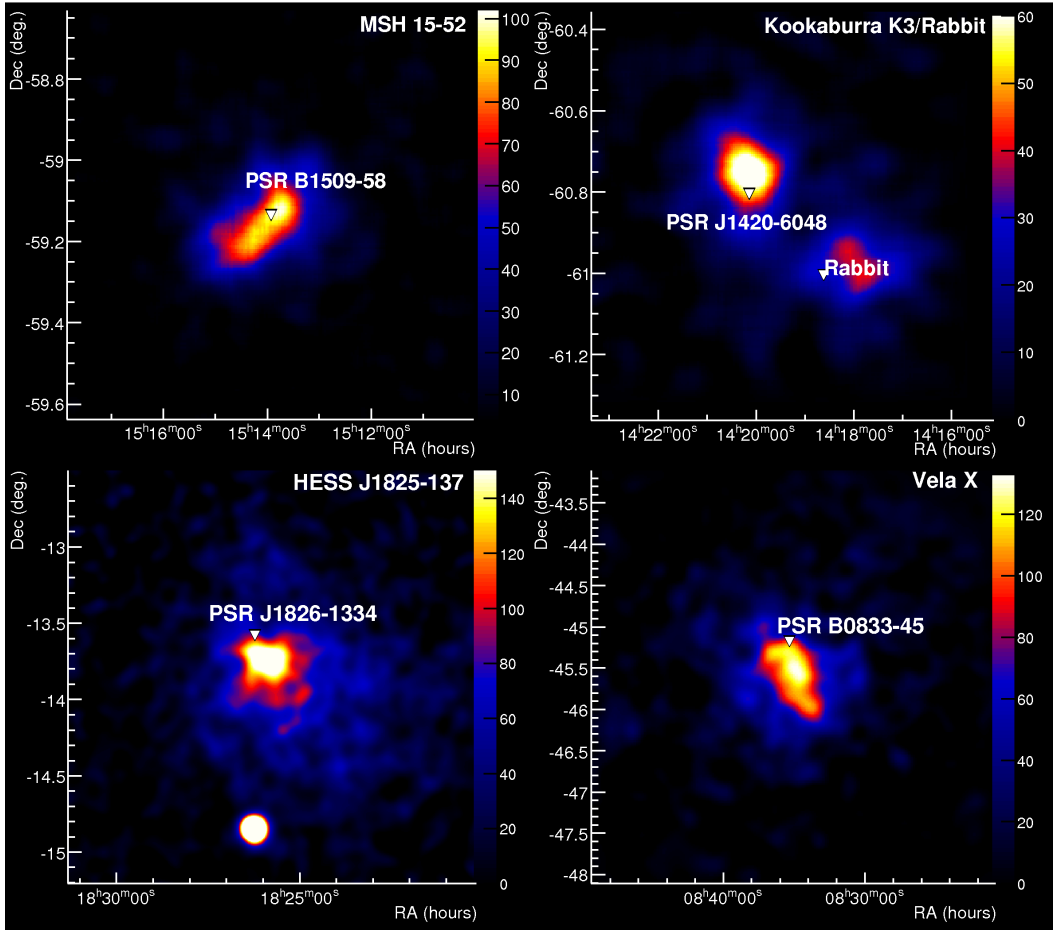}
  \caption{Acceptance-corrected smoothed excess maps of
  MSH--15--5\emph{2} (top left)~\citep{HESSMSH}, the Kookaburra region
  showing the $\gamma$-ray emission coincident with the two
  non-thermal wings of the Kookaburra (top
  right)~\citep{HESSKookaburra}, HESS\,J1825--137 (bottom
  left)~\citep{HESS1825, HESS1825_II} and Vela~X (bottom
  right)~\citep{HESSVelaX}. Also shown are the energetic pulsars that
  are thought to power the PWNe.}
  \label{fig::pwn_morphology}
\end{figure*}

\begin{figure*}
  \centering
  \includegraphics[width=0.9\textwidth]{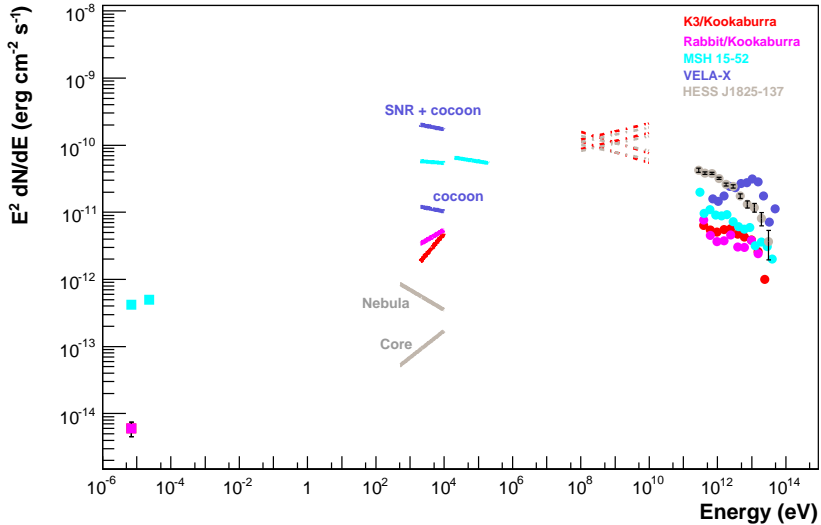}
  \caption{Spectral energy distribution for the PWNe in the Kookaburra
  region (K3 and Rabbit) (red), MSH--15-5{\emph2} (turquoise), Vela~X
  (blue), and HESS\,J1825--137 (grey). The similar energy flux for the
  $\gamma$-ray emission in comparison to the vastly different energy
  flux for the X-ray emission is apparent.}
  \label{fig::pwn_spectra}
\end{figure*}

Apart from the Crab Nebula, no individual PWNe have been unambiguously
associated with EGRET sources, although several unidentified EGRET
sources are located in close proximity to prominent PWN, such as in
the Kookaburra region, or MSH--15--5\emph{2}. PWN are however one
candidate for the population of slowly varying low-latitude
unidentified sources. GLAST will shed more light on this population
and possibly establish PWN as emitters in the MeV to GeV range. In VHE
$\gamma$-rays PWN make up the majority of the identified Galactic
sources detected so far~\citep{FunkBarcelona, YvesBarcelona}. Apart
from the Crab Nebula (the brightest steady VHE $\gamma$-ray source)
several prominent PWN were identified in VHE $\gamma$-rays in the last
two years. These detections include
MSH--15--5\emph{2}~\citep{HESSMSH}, Vela~X~\citep{HESSVelaX}, the two
sources in the Kookaburra region~\citep{HESSKookaburra} and lately
HESS\,J1825--137~\citep{HESS1825, HESS1825_II}. VHE $\gamma$-ray
emission from PWN comes in various disguises as shown in
Figure~\ref{fig::pwn_morphology}: These include a) point-like emission
such as from the Crab Nebula~\citep{HESSCrab} and from the composite
SNR G\,0.9+0.1, where the $\gamma$-ray emission was shown to originate
from the central PWN~\citep{HESSG0.9}, b) emission tracing the X-ray
contours around a central pulsar such as in MSH--15-5\emph{2} or c)
asymmetrically extending to one side and tracing the X-ray contours
such as in Vela~X and finally d) the emerging new class of offset PWN
exemplified by HESS\,J1825--137 where the $\gamma$-ray emission shows
a similar morphology to the X-ray emission but on a much larger
scale~\citep{HESS1825_II}. Calculating the efficiency of the energetic
pulsars powering the PWN that is necessary to account for the VHE
$\gamma$-ray luminosity, values between 0.02\% (Crab Nebula) and 7.5\%
(HESS\,J1825--137) of the spin-down luminosity are found. The
broadband SEDs of the VHE $\gamma$-ray PWN can typically be well
described by leptonic models, although claims have been made for a
hadronic component at the high-energy end of the
spectrum~\citep{HornsHadronicPWN}. Vela~X is the first VHE
$\gamma$-ray source in which the peak in the Inverse Compton energy
flux has been detected within the H.E.S.S.\ energy
range~\citep{HESSVelaX}. Figure~\ref{fig::pwn_spectra} shows, that
while the ranges of $\gamma$-ray fluxes for the detected PWN is rather
small, the differences in the X-ray energy fluxes span a large
range. This might be alluded to largely different magnetic fields, to
different angular scales on which the X-ray emission has been measured
or simply to the fact, that different populations of electrons are
responsible for the X-ray and the $\gamma$-ray emission.

The most prominent example of the new class of {\emph{offset PWN}} is
HESS\,J1825--137. This object can serve as a template for a whole new
class of $\gamma$-ray PWN in which a) the $\gamma$-ray emission is
shifted away from the pulsar, possibly due to dense material on one
side that prevents an isotropic expansion of the PWN and b) the size
of the VHE $\gamma$-ray PWN is on a much larger scale ($\sim
1^{\circ}$) than the X-ray PWN ($\sim 1'$)~\citep{XMM1825}.
Concerning the offset morphology, asymmetric reverse shock
interactions were first proposed to explain the offset morphology of
the Vela~X PWN based on hydro-dynamical simulations
by~\citet{Blondin}. The different sizes for the $\gamma$-ray and X-ray
PWNe can be explained by the difference in the synchrotron cooling
lifetimes of the (higher energy) X-ray emitting and the (lower energy)
IC-$\gamma$-ray emitting electrons. The $\gamma$-ray sources that can
be explained in this framework are typically extended, their emission
region overlaps with energetic pulsars (energetic enough to explain
the $\gamma$-ray flux by their spindown power) and very importantly
also show evidence for an X-ray PWN. So far only Vela~X and
HESS\,J1825--137 match this picture, several other unidentified VHE
$\gamma$-ray sources have been proposed to be offset PWN, but all
these cases lack the detection of an X-ray PWN.

\section{New Supernova remnants found in VHE $\gamma$-rays} 
\label{sec::HESS_new}
Originally Supernova remnants have been detected by means of radio
observation sensitive to synchrotron emission in magnetic
fields. Radio observations are particularly suited to detect Supernova
remnants in the inner Galaxy since they are insensitive to the
prevailing dust emission within the Galactic plane. The same holds for
hard X-ray emission and is particularly true for $\gamma$-rays. A
survey of the Galaxy in the $\gamma$-ray regime proved to be a good
means to detect new $\gamma$-ray SNRs, that are inconspicuous in other
wavebands. H.E.S.S.\ observations of the inner part of the Galactic
plane revealed $\sim 20$ new $\gamma$-ray sources~\citep{HESSScan,
HESSScanII, FunkBarcelona}. While some could be identified at other
wavebands, such as the Supernova remnants described in the previous
sections or the microquasar LS\,5039~\citep{HESSLS5039,
HESSLS5039_II}, most objects were left unidentified following their
detection. A programme of detailed MWL studies using existing radio
and X-ray facilities is underway to establish positional counterparts
of the $\gamma$-ray sources.

\begin{figure*}
  \centering
  \includegraphics[width=0.99\textwidth]{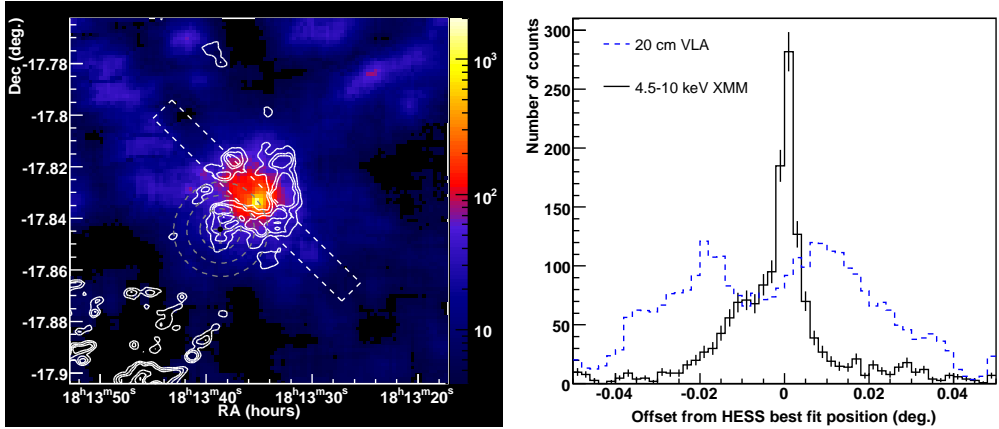}
  \caption{Comparison of radio, and X-ray data of
  HESS\,J1813--178. {\bf Left:} XMM-Newton counts map above 4.5~keV of
  the region surrounding HESS\,J1813--178 (colour contours) smoothed
  with a Gaussian kernel of width 0.002$^{\circ}$. The extended tail
  towards the north-east is visible in this figure. Overlaid is the
  20~cm shell-like emission (white contours) as detected by the
  VLA~\citep{Brogan1813}. Also shown are the positional contours of
  the best fit position of HESS\,J1813--178 (dashed circles correspond
  to the 1, 2, and 3$\sigma$ positional confidence contours) as given
  in~\citet{HESSScanII}. {\bf Right:} Slice through the emission in
  radio and X-rays as plotted on the left hand side. The box in which
  the slices were determined is also given in the left panel (white
  box). The X-ray slice shows the compact core with the slice towards
  the north-east, whereas the radio slice shows the shell-like
  structure of the emission.}
  \label{fig::new_1813}
\end{figure*}

A particularly interesting object found in the survey of the Galactic
plane is HESS\,J1813--178. At first flagged
unidentified~\citep{HESSScan}, it was quickly found to be positionally
coincident with: a) A previously unpublished archival faint radio
(VLA) source (20 cm) showing a shell-like
morphology~\citep{Brogan1813} b) a previously unpublished archival
bright X-ray ASCA source (2--10 keV)~\citep{Brogan1813}, and c) a hard
X-ray INTEGRAL source 10--100 keV~\citep{Ubertini1813}. H.E.S.S., ASCA
as well as INTEGRAL lacked the spatial resolution to resolve the
object. The VLA radio source showed a shell-like morphology,
suggesting that the VHE $\gamma$-ray emission originates in the shell
of a Supernova Remnant. However, in a 30~ksec XMM-Newton X-ray
observation non-thermal synchrotron emission was found not from a
shell, but rather from an object embedded within the shell with a
faint tail towards the north-east (see Figure~\ref{fig::new_1813})
resembling in its shape a PWN~\citep{Funk1813}. This detection reveals
that HESS\,J1813--178 is connected to a composite SNR similar to
e.g. the $\gamma$-ray source G0.9+0.1~\citep{HESSG0.9}. In
HESS\,J1813--178 for the first time, an SNR initially detected in VHE
$\gamma$-rays and subsequently confirmed with superior angular
resolution radio and X-ray data. This detection shows, that
$\gamma$-ray observations especially in the Galactic plane are well
suited to detect SNR, that are otherwise hard to detect due to
obscuration and dust absorption. Other objects tentatively connected
to shell-type or composite SNRs are
HESS\,J1640--465~\citep{HESSScanII, Funk1640}, and
HESS\,J1834--087~\citep{HESSScanII, FunkBarcelona}.

\section{Summary and Prospects for GLAST}
\label{sec::glast}
The upcoming GLAST satellite, in its energy range between 30~MeV and
300~GeV the successor to EGRET will provide a unique tool for studying
Supernova remnants. Especially in crowded regions in the Galactic
plane dominated by diffuse emission, the improved angular resolution
of the instrument in comparison to EGRET will be important for
disentangling source confusion and identification of counterparts. As
obvious from Figure~\ref{fig::shell_spectra}, GLAST will provide
spectral measurements in an energy regime in which differences between
hadronic and leptonic production $\gamma$-ray production mechanisms
are significant. Thus GLAST might finally disentangle whether the
$\gamma$-ray emission detected by H.E.S.S.\ is in fact the first
direct evidence of accelerated hadrons in shell-type SNRs, a question
that directly relates to the origin of cosmic rays. GLAST will provide
a highly even sky-coverage on long timescales and therefore population
studies of SNRs will be possible. Since EGRET is thought to have just
not been sensitive enough to single out individual PWNe and enable us
to conclude about their population, it is expected that these objects
will appear as a source classin the GLAST sky. Population studies of
SNRs provides information not only on individual objects but on
spatial-statistical aspects that can be used to understand the
transition of source populations through the regime of GeV-cutoffs as
already evident in numerous EGRET sources. Already now SNRs are an
established source class in VHE $\gamma$-ray astronomy. Observations
in this wavebands provide an important tool to understand the
acceleration processes within these Galactic accelerators.


\section{Acknowledgements}
The author would like to acknowledge the support of their host
institutions, and additionally support from the German Ministry for
Education and Research (BMBF) and from the Department of energy (DOE).
The author would like to thank the whole H.E.S.S.\ collaboration for
their support, COSPAR for the financial support, the unknown
referee(s) for their valuable comments, and finally the organisers of
the session E1.4 J.Vink and P.~O. Slane for the invitation to the
conference.


\end{document}